\documentclass[%
 reprint,
superscriptaddress,
 amsmath,amssymb,
 aps,
]{revtex4-2}

\usepackage{graphicx}
\usepackage{dcolumn}
\usepackage{bm}
\usepackage{color,amsmath}
\usepackage{physics}
\usepackage{siunitx} 
\usepackage{hyperref}

\newcommand{\TUWien}{Vienna Center for Quantum Science and Technology, Atominstitut, TU Wien, Stadionallee 2, 1020 Vienna, Austria}
\newcommand{\be}{\begin{equation}}
\newcommand{\ee}{\end{equation}}

\newcommand{\god}{g_\mathrm{1D}}
\newcommand{\nod}{n_\mathrm{1D}}
\newcommand{\delrho}{\delta\rho}

\newcommand{\chg}[1]{\textcolor{black}{ #1}}

\begin{document}

\title{Experimental verification of the area law of mutual information in a quantum field simulator}

\author{Mohammadamin~Tajik}
\email{amintajik.physics@gmail.com}
\affiliation{\TUWien}%
\author{Ivan~Kukuljan}
\affiliation{Max-Planck Institute of Quantum Optics, Hans-Kopfermann-Str. 1, 85748 Garching, Germany}%
\affiliation{Munich Center for Quantum Science and Technology (MCQST), Schellingstr. 4, D-80799 M{\"u}nchen}%
\author{Spyros~Sotiriadis}
\affiliation{Dahlem Center for Complex Quantum Systems, Freie Universit{\"a}t Berlin, Berlin, Germany}
\author{Bernhard~Rauer}
\affiliation{\TUWien}
\affiliation{Laboratoire Kastler Brossel, ENS-Universit\'{e} PSL, CNRS, Sorbonne Universit\'{e}, Coll\'{e}ge de France, 24 rue Lhomond, 75005 Paris, France}
\author{Thomas~Schweigler}
\affiliation{\TUWien}
\author{Federica~Cataldini}
\affiliation{\TUWien}
\author{Jo{\~a}o~Sabino}
\affiliation{\TUWien}
\affiliation{Instituto de Telecomunica\c{c}\~{o}es, Physics of Information and Quantum Technologies Group, Av. Rovisco Pais 1, 1049-001, Lisbon, Portugal}
\affiliation{Instituto Superior T\'{e}cnico, Universidade de Lisboa, Av. Rovisco Pais 1, 1049-001, Lisbon, Portugal}
\author{Frederik M{\o}ller}
\author{Philipp~Sch{\"u}ttelkopf}
\author{Si-Cong~Ji}
\affiliation{\TUWien}
\author{Dries~Sels}
\affiliation{Department of physics, New York University, New York City, NY 10003, USA}
\affiliation{Center for Computational Quantum Physics, Flatiron Institute, 162 5th Ave, New York, NY 10010, USA}
\author{Eugene~Demler}
\affiliation{Institute for Theoretical Physics, ETH Zurich, Wolfgang-Pauli-Str. 27, 8093 Zurich, Switzerland}
\author{J{\"o}rg~Schmiedmayer}
\email{schmiedmayer@atomchip.org}
\affiliation{\TUWien}

\date{\today}

\begin{abstract}
Theoretical understanding of the scaling of entropies and the mutual information has led to significant advances in the research of correlated states of matter, quantum field theory, and gravity.
Measuring von Neumann entropy in quantum many-body systems is challenging as it requires complete knowledge of the density matrix.
In this work, we measure the von Neumann entropy of spatially extended subsystems in an ultra-cold atom simulator of one-dimensional quantum field theories. We experimentally verify one of the fundamental properties of equilibrium states of gapped quantum many-body systems, the area law of quantum mutual information. We also study the dependence of mutual information on temperature and the separation between the subsystems. Our work is a crucial step toward employing ultra-cold atom simulators to probe entanglement in quantum field theories.
\end{abstract}

\maketitle

The study of quantum information measures is central to a wide range of areas in physics, from condensed matter and atomic physics to high energy physics and gravity~\cite{HorodeckiReview2009,OrusTN2014,CiracApproximability2008,AbaninMBL2019,SenthilSymmetry2015, Wilczek1,Wilczek2,Srednicki,Calabrese-Cardy_review_2004,Calabrese-Cardy_review_2009,polchinski_black_2017,ryu_holographic_2006}.
Some of the most commonly studied quantities in quantum information are the (entanglement) entropy and the quantum mutual information. If a system described by the density matrix $\varrho$ is composed of subsystems $A$ and $B$, then the von Neumann (vN) entropy of subsystem $A$ is defined as
\be
S_A=-\text{Tr}\left(\varrho_A \ln \left(\varrho_A \right) \right)\, ,
\ee
where $\varrho_A=\text{Tr}_B \left(\varrho \right)$ is the reduced density matrix of subsystem $A$. If the state $\varrho$ is pure ($\text{Tr}\left( \varrho^2 \right)=1$), then the vN entropy is a measure of entanglement between $A$ and $B$, thus called entanglement entropy, where $S_A=S_B$. For mixed states ($\text{Tr}\left(\varrho^2 \right)<1$), the vN entropy captures both classical and quantum correlations, and it is no longer a good measure of entanglement. For mixed states, several other measures and witnesses of entanglement have been studied, with the positivity of partial transpose criterion \chg{and quantum discord being prominent examples~\cite{HorodeckiReview2009, vedral2012quantumdiscord}}. In cases where, rather than entanglement, the shared amount of information between two subsystems $A$ and $B$ is of interest, the quantum mutual information (MI),
\be
I(A:B)=S_A+S_B-S_{A\cup B} \, , \label{eq:MI}
\ee
is a central object of study. It measures the total amount of correlation between the two subsystems, including all higher-order correlations for both pure and mixed states. For pure states, the value of the MI is equal to twice the entanglement entropy of one of the subsystems. 

\begin{figure*}[t]
    \centering
    \includegraphics[width=\linewidth]{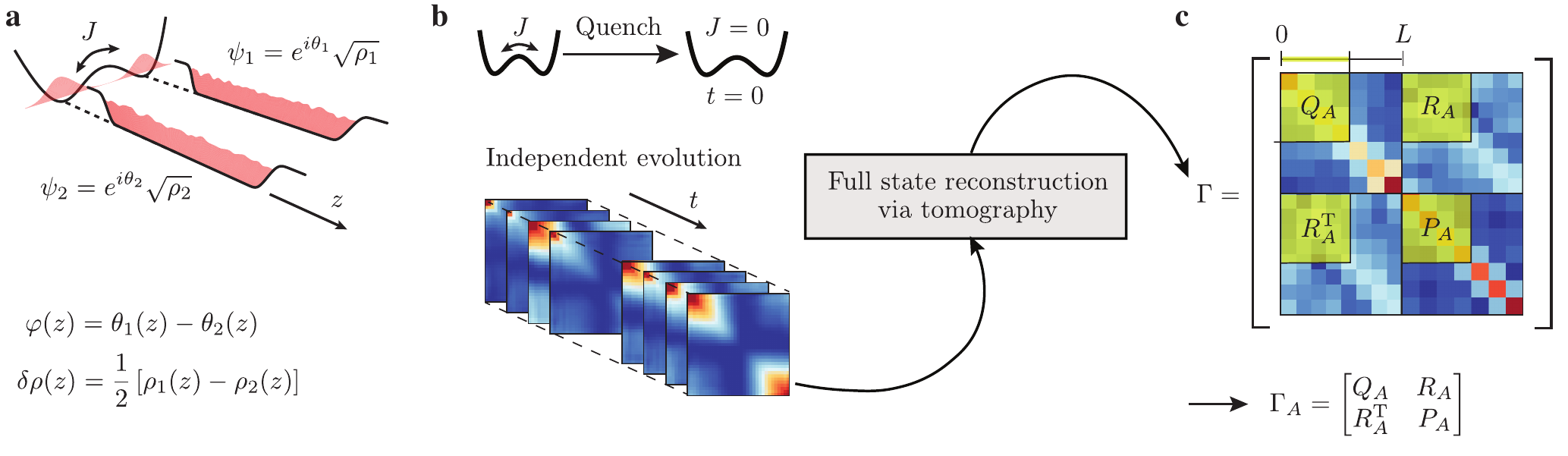}
    \caption{\textbf{Schematic of the experimental protocol.} \textbf{a}, The experimental protocol starts by cooling down a pair of tunnelling coupled superfluids in a double-well potential with a finite single particle tunnelling rate $J$, typically $\approx 2\pi \times 1\, \si{\hertz}$. The radial trapping frequency is $\omega_\perp = 2\pi \times 1.4\, \si{\kilo \hertz}$ and typical linear densities are $\nod \approx 70\, \si{\micro \meter^{-1}}$. \textbf{b}, Right after the cooling process, the tunnelling rate is changed to zero in $2\, \si{\milli \s}$. At $t=0$, the two condensates are already uncoupled and their independent evolution following the Tomonaga-Luttinger liquid Hamiltonian starts. The measured phase correlations at different times are used to fit the full covariance matrix $\Gamma$. \textbf{c}, To calculate the vN Entropy for a subsystem, $S(\Gamma_A)$, we use the correlations in that specific region (yellow shaded area). \chg{Note that $Q$, $R$, and $P$ have different units and the range of the color axis are different. All tomography results are presented and discussed in SM.}}
    \label{fig:experiment}
\end{figure*}

Information theory measures reveal one of the fundamental properties of quantum many-body systems, the area laws~\cite{CiracApproximability2008,Eisert2010}. It was first noticed in gravitational physics, that surprisingly, the entropy of a black hole is proportional to the surface area of its event horizon and not its volume \cite{polchinski_black_2017}. Interestingly, a similar property was found in quantum many-body systems: The vN entropy of ground states of systems with a gapped Hamiltonian scales with the surface area of the subsystem and not its volume \cite{Eisert2010}. Even more general, thermal states of systems with a gapped Hamiltonian exhibit an area law of mutual information \cite{WolfCiracMI2008}. This means that the information shared between parts of a quantum many-body system is only considerable over a short distance, set by the correlation length. In particular, such a bound on the required information to model a many-body system provides the foundations for the overwhelming success of tensor network-based methods \cite{OrusTN2014}. 
In contrast, it is known that critical systems described by conformal field theory exhibit a logarithmic scaling instead of the area law \cite{Calabrese-Cardy_review_2004}. In thermal states, while the mutual information has an area or a log law, the vN entropy will exhibit a volume law where it is proportional to the volume of the subsystem.

Extracting quantum information measures in quantum many-body systems has been the aim of several experiments~\cite{LuoEntropy2007,GreinerEntanglement2015, BlochEntropy2015,CocchiEntropy2017, ZollerEntanglement2019}. Calculating the vN entropy requires access to the density matrix of the full system, $\varrho$, which usually \chg{requires a full state tomography in different experimental platforms. Instead of $\varrho$, several techniques are developed to measure the purity, $\text{Tr} \left(\varrho^2 \right)$, which enables the calculation of second-order Rényi entropy, $S_2 = -\ln(\text{Tr} \left(\varrho^2 \right) )$. These methods can be based on the interference of two identical copies of a quantum system~\cite{GreinerEntanglement2015} or randomised measurements on a single copy~\cite{ZollerEntanglement2019}. Note that, in these examples, the purity is directly measured but the full state is not reconstructed, hence the calculation of vN entropy is not feasible.}

\begin{figure}[t]
    \centering
    \includegraphics[width=\linewidth]{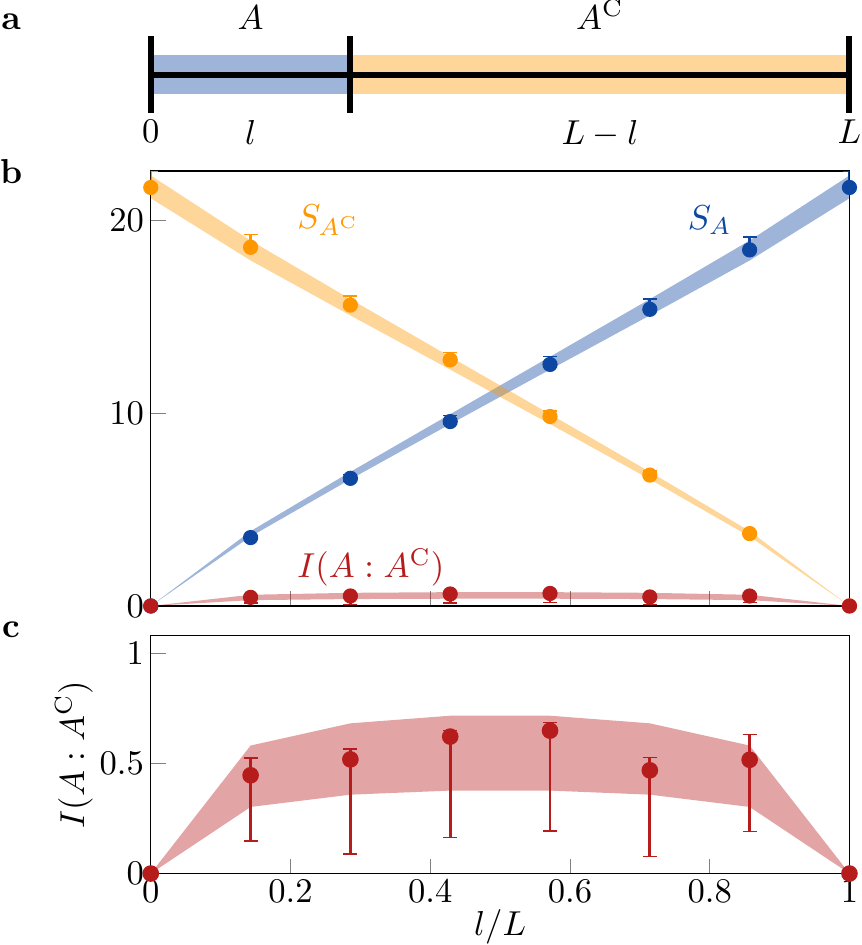}
    \caption{\textbf{Area law of MI and volume law of vN entropy.} \textbf{a}, A system of size $L=49\,\si{\micro \meter}$ is divided into a subsystem $A$ of length $l$ and its complement subsystem of length $(L-l)$. \textbf{b}, The experimental results for $I(A:A^\mathrm{C})$, $S_A$, and $S_{A^\mathrm{C}}$, calculated based on $N=7$ modes, are plotted as bullets with error bars representing the $95\%$ confidence intervals obtained via bootstrapping~\cite{efron1986bootstrap}. The shaded areas show the $95\%$ confidence interval for the theory predictions, considering the uncertainty in the estimated temperature and tunnelling rate $J$. \textbf{c}, Close up of the measured MI.}
    \label{fig:s_mi_l}
\end{figure}

Nevertheless, several optical lattice setups are able to measure the vN entropy, which is obtained either from a single site reduced density matrix or as a classical thermodynamic entropy of the whole system~\cite{LuoEntropy2007,CocchiEntropy2017}. In special cases, when the system is diagonal enough, even many-body vN entropy can be accessed \cite{evrard_observation_2021}.
However, the measurement of vN entropy between extended spatial subsystems has so far remained elusive, as well as the verification of the predicted area law scaling of the MI.

In this work, we address these challenges and study the scaling of the vN entropy and the MI with subsystem size in a continuous quantum many-body system. Our setup is a pair of tunnelling-coupled quasi-one-dimensional (1D) ultracold Bose gases (see Fig. \ref{fig:experiment}\textbf{a}), cooled down and trapped below an atom chip~\cite{folman2000firstatomchip}.
Along the longitudinal axis, $z$, the clouds are confined in box-like potentials with hard walls, created by superposing magnetic and optical dipole potentials~\cite{tajik2019designing}. In one of the transverse directions, the atoms are trapped in a double-well potential, created by dressing with radiofrequency fields. The single-particle tunnelling rate $J$ between the two condensates is adjusted by changing the amplitude of the radiofrequency fields~\cite{schumm2005matterwave}.

\begin{figure}[t]
    \centering
    \includegraphics[]{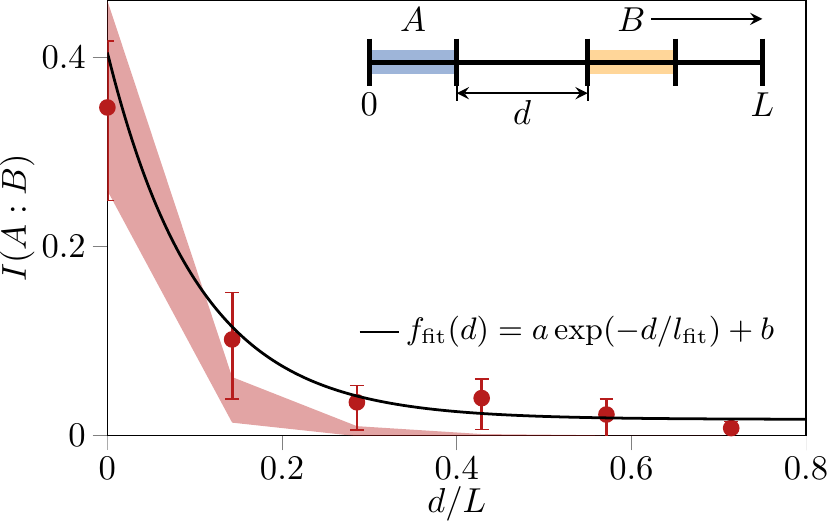}
    \caption{\textbf{Shared information content between two spatially separated subsystems.} Here, we calculate the MI of two disjoint subsystems with the same length, $l/L = 0.15$, as a function of the distance between them, $d$. As demonstrated in the inset, while the subsystem $A$ is kept fixed on the left edge of the system, $B$ is shifted away to the other edge. Bullets and shadings represent experimental and theoretical data for $I(A:B)$ respectively (see Fig. 2 for more details on the error bars and shaded area). The solid black line is an exponential fit with $l_\mathrm{fit} = 5.1\, \si{\micro \meter}$. The finite bias $b$ results from estimating a positive quantity, MI, using finite statistics.}
    \label{fig:mi_gap}
\end{figure}

To prepare a state in thermal equilibrium, we directly cool down an atomic cloud of $^{87}$Rb in a strongly coupled double-well, using standard techniques of laser cooling and evaporative cooling ~\cite{rauer2006coolingoned}.
The quantum fields describing each condensate can be written in phase-density representation as $\psi_\mathrm{n}(z) = \exp(i\theta_\mathrm{n})\sqrt{\rho_\mathrm{n}}$, with $\mathrm{n} = 1,2$. The spatially resolved relative phase between the two condensates, $\varphi(z) = \theta_1(z) -\theta_2(z)$, is extracted from interference images, taken $15.6\, \si{\milli \s}$ after releasing the atoms from the trap and letting them fall freely. In the limit of low energy excitations, the correlations of the relative phase are well described by the sine-Gordon Hamiltonian~\cite{exp_sG}. Expanding the interaction term in the case of high tunnelling rates leads to the massive Klein-Gordon Hamiltonian,
\begin{multline}\label{eq:KG}
    H_\mathrm{KG} = \int_0^L \mathrm{d}z \biggl[\god \delrho^{2}(z) + \frac{\hbar^2 \nod}{4m}\left(\partial_{z}\varphi(z)\right)^{2} \\  + \hbar J \nod \varphi^2(z)\biggr]\, ,
\end{multline}
where $L$ is the length of a system with uniform averaged 1D density $\nod$, $\god$ is the 1D interaction strength, $m$ is the mass of an atom, and the relative density, $\delrho(z)=\left[\rho_1(z)-\rho_2(z)\right]/2$, is the conjugate field of the relative phase, fulfilling $[\varphi(z),\delrho(z^\prime)] = i\mathrm{\delta}(z-z^\prime)$ . A direct measurement of $\delrho(z)$ is unfeasible in our current experimental setup. Hence, to reconstruct the full covariance matrix of the initial state, 
\begin{equation}
    \Gamma=\mqty[ Q & R\\ R^\mathrm{T} & P]\, ,
\end{equation}
we use a tomography procedure~\cite{Tomography}. Here, $Q_{ij}=\left<\varphi(z_i)\varphi(z_j)\right>$, $P_{ij}=\left<\delrho(z_i)\delrho(z_j)\right>$, $R_{ij}=\left<\frac{1}{2}\left\{\varphi(z_i),\delrho(z_j)\right\}\right>$ with $i,j\in\{1,\ldots,N\}$, where $z_i$ denotes different points on a discrete grid with $N$ points. The upper limit for $N$ is given by the resolution of the imaging system, which limits our access to higher momentum modes (larger than $N$) and enforces an ultra-violet (UV) cut-off.

To apply this tomographic method, we ramp up the barrier between the two strongly coupled condensates in $2\, \si{\milli \s}$, and let them evolve independently in the uncoupled double-well ($J=0$). \chg{We then directly measure the phase-phase correlations for different evolution times after the quench.} The post-quench dynamics follow the Tomonaga-Luttinger liquid Hamiltonian. Over time, the initial eigenmodes of the relative density rotate into the phase quadrature \chg{and vice versa}, enabling us to access the information about these eigenmodes by fitting the initial second-order correlation functions of phase-density and density-density to the measured evolution of the phase-phase correlations in the momentum space. A thorough explanation of the reconstruction process and its results is given in SM.

\chg{In the experiment, we prepare initial states that are thermal equilibrium states of Klein-Gordon Hamiltonian. The quadratic form of this Hamiltonian, as shown in (\ref{eq:KG}), implies that the prepared initial states are Gaussian. After a fast quench of $J$ to zero, the initial state evolves under another quadratic Hamiltonian, ensuring that the state remains Gaussian under evolution ~\cite{Gaussification1}.} To confirm \chg{Gaussianity}, we measure the normalized averaged connected fourth-order correlation function, $M^{(4)}$\chg{, and show that the higher order correlations are negligible. Note that the tomography process does not include any restrictions regarding the Gaussianity of the reconstructed state.}

Having Gaussian states significantly simplifies the calculation of the quantum information measures - an otherwise highly non-trivial task in quantum field theory \cite{serafini_entanglement_2017}. Gaussian states are fully described by their covariance matrix $\Gamma$.
Its symplectic spectrum is obtained by diagonalizing $i\mathcal{J}\Gamma$ where $\mathcal{J}=\mqty[ 0 & I\\ -I & 0]$ is the symplectic unit. The symplectic spectrum consists of pairs $\pm\gamma_n$, $n\in\{1,\ldots,N\}$. It encodes the complete information that is contained in the covariance matrix~\chg{\cite{Eisert2010}}. Consequently, it can be used to reconstruct the full density matrix of the state and the measures of quantum information. In particular, the vN entropy is given by
\begin{multline}
    S(\Gamma)=\sum_{n=1}^{N}\left[\left(\gamma_{n}+\tfrac{1}{2}\right)\ln\left(\gamma_{n}+\tfrac{1}{2}\right)\right. \\
    \left.-\left(\gamma_{n}-\tfrac{1}{2}\right)\ln\left(\gamma_{n}-\tfrac{1}{2}\right)\right] \label{eq:vNentropyGamma}
\end{multline}
For non-Gaussian states, neglecting higher-order correlations and estimating the vN entropy based on the covariance matrix gives a lower bound to the actual entropy~\cite{Eisert-Wolf_2007}.

\begin{figure}[t]
    \centering
    \includegraphics[width =\linewidth]{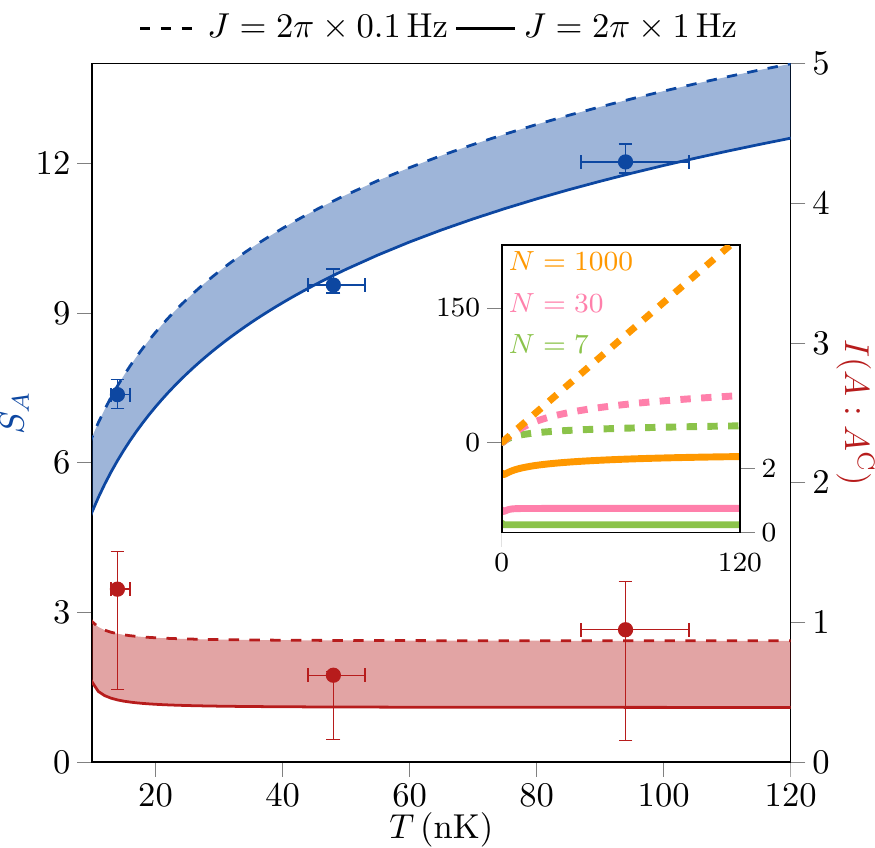}
    \caption{\textbf{Temperature dependence of MI and vN entropy.} The bullets represent experimental data for $S_A$ (left axis) and $I(A:A^\mathrm{C})$ (right axis) with $N = 7$ for measurements with three different temperatures, but otherwise similar parameters. The size of the subsystem $A$ is $l/L = 0.4$ (see Fig. \ref{fig:s_mi_l}\textbf{a}). The error bars show the $95\%$ confidence intervals achieved via bootstrapping. The shaded areas represent the calculations based on the theoretical model for different tunnelling rates, between $J = 2\pi \times 0.1\,\si{\Hz}$ (dashed lines) and $2\pi \times 1\,\si{\Hz}$ (solid lines). The inset shows the theoretical predictions for $S_A$ (left axis) and $I(A:A^\mathrm{C})$ (right axis) for three different values of $N$, for $J = 2\pi \times 0.4\,\si{\Hz}$ without considering the effect of finite optical resolution (see SM).} 
    \label{fig:s_mi_temp}
\end{figure}

Having the reconstructed initial covariance matrix at hand, we use expression (\ref{eq:vNentropyGamma}) to calculate the vN entropy of any subsystem $A$, $S_A(\Gamma_A)$ (see Fig. \ref{fig:experiment}\textbf{c}). Using equation (\ref{eq:MI}), the MI between two subsystems $A$ and $B$ is calculated. To observe the scaling of the vN entropy and the MI, we calculate the vN entropy of subsystems with different lengths and consecutively the MI with their complement subsystems as illustrated in Fig.~\ref{fig:s_mi_l}\textbf{a}. As expected for thermal states, the vN entropy is in the volume law regime, depending linearly on the size of the subsystem (Fig. \ref{fig:s_mi_l}\textbf{b}). Measuring the vN entropy allows us to study the scaling of the mutual information. We find an area law for MI, with a plateau forming in the bulk of the system (Fig. \ref{fig:s_mi_l}\textbf{c}). Our results represent an experimental verification of one of the elementary features of quantum many-body systems \cite{WolfCiracMI2008}. 

We continue by studying the dependence of MI on the distance between two subsystems. In this case, we calculate the MI of two subsystems $A$ and $B$, separated by a gap of length $d$. The results are presented in Fig. \ref{fig:mi_gap}. As expected, the MI decreases as the two subsystems get further apart. We can extract a decay length by fitting an exponential function to the experimental data.  The fitted decay length, $l_\mathrm{fit} = 5.1\, (3.7,\, 8)\, \si{\micro \meter}$ agrees with the correlation length $l_\mathrm{C} = 6.8\, (6,\, 7.7)\, \si{\micro \meter}$ calculated based on the experimental parameters (see SM). The intervals represent the $95\%$ confidence intervals obtained via bootstrapping. 

For typical temperatures of our experiment, the vN entropy depends linearly on the temperature. The linear dependence changes to a logarithmic dependence by introducing a finite UV cut-off, as shown in the inset of Fig. \ref{fig:s_mi_temp} (dashed lines). As temperature increases, the symplectic eigenvalues grow, and the calculated entropy using a finite number of modes saturates. The MI, however, regardless of the UV cut-off, reaches a finite asymptotic value, given by the classical correlations \cite{Katsinis2020}. Reducing the number of modes reduces the asymptotic value due to the limited available information in the modes taken into account, as presented in the inset of Fig. \ref{fig:s_mi_temp} (solid lines). In Fig. \ref{fig:s_mi_temp}, measured vN entropy and MI for three different temperatures are illustrated along with the theoretical predictions. The measurements agree with the theory calculated for the extracted parameters and $N=7$ lowest modes.

It is important to stress that the only assumption we make to calculate the vN entropy and the MI is that the post-quench dynamics follow a Tomonaga-Luttinger liquid Hamiltonian, which has been confirmed in the previous work~\cite{Gaussification1}. Our measurements do not rely on any assumption related to the Gaussianity of the initial state. We rather confirm that the initial state is Gaussian by measuring the higher-order correlations. Even for non-Gaussian initial states, our results would represent a lower bound to the entropy of the full state. 

The results presented here are a step towards the more ambitious goal of measuring many-body entanglement in a continuous 1D quantum system. Reconstruction of the full covariance matrix enables us to calculate any entanglement measure applicable to both pure and mixed states, such as logarithmic negativity. However, there are two main limiting factors preventing us from detecting the entanglement. The entanglement can only be detected if a sufficient number of momentum modes are measured whose mode occupation numbers are close to the ground state. In our current experiments, the nonzero temperature ($10\,$-$\,100\, \si{\nano \K}$) of the Bose gas keeps the occupation numbers of the lower momentum modes too high above the value of the ground state. At the same time, the finite optical resolution introduces a soft cut-off which not only prevents us from measuring higher momentum modes but also modifies the lower modes that can be measured. Improving any of these two aspects would make the measurement of the entanglement possible. 

Another promising direction for future work is to go beyond quadratic models and detect entanglement in an interacting model. It has already been demonstrated that the atom chip experiments can successfully simulate the sine-Gordon model, and higher-order correlation functions can be measured~\cite{exp_sG}. Developing a tomography procedure for this setting would give us access to entanglement properties in interacting many-body quantum systems~\cite{KSTcorrelations2018,KukuljanHTentanglement2022}.

\begin{figure*}[t]
    \centering
    \includegraphics[width = 0.8\linewidth]{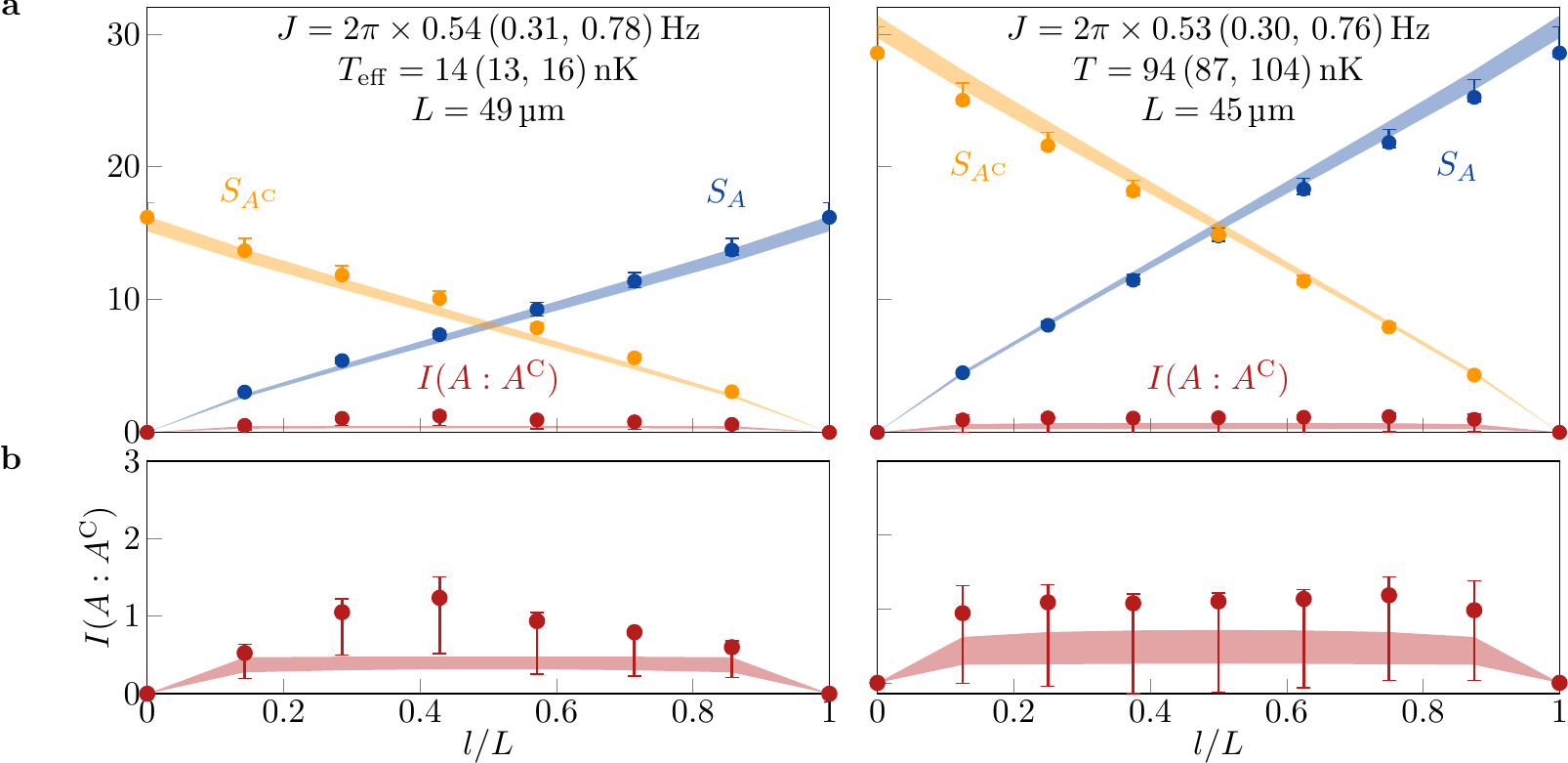}
    \caption{\textbf{Additional results for area law of MI and volume law of vN entropy.}  \textbf{a}, Experimental results for $I(A:A^\mathrm{C})$, $S_A$, and $S_{A^\mathrm{C}}$, calculated based on $N=7$ modes (left) and $N=8$ modes (right) (see Fig. \ref{fig:s_mi_l} for detailed explanation). The extracted parameters including their $95\%$ confidence intervals in parenthesis are given above. \textbf{b}, Close up of the measured MI.}
    \label{fig:extnd_s_mi}
\end{figure*}

\begin{acknowledgments}

We want to thank M. Gluza, J. Eisert, I. Cirac, I. Mazets and S. Erne for their helpful discussions. This work is supported by the DFG/FWF Research Unit FOR 2724 ‘Thermal machines in the quantum world’, the FQXi program on ‘Information as fuel’. The Flatiron Institute is a division of the Simons Foundation. ED acknowledges support from the ARO grant number W911NF-20-1-0163. The work of I.K. was supported by the Max-Planck-Harvard Research Center for Quantum Optics (MPHQ). F.C., F.M., and J. Sabino acknowledge support from the Austrian Science Fund (FWF) in the framework of the Doctoral School on Complex Quantum Systems (CoQuS). J. Sabino acknowledges support by the Funda\c{c}{\~a}o para a Ci\^{e}ncia e Tecnologia, Portugal (PD/BD/128641/2017). T.S. acknowledges support from the Max Kade Foundation through a postdoctoral fellowship. B.R. acknowledges support by the European Union’s Horizon 2020 research and innovation programme under the Marie Sk\l{}odowska-Curie grant agreement No. 888707. D.S. is partially supported by AFOSR: Grant FA9550-21-1-0236. S.S. acknowledges support by the European Union’s Horizon 2020 research and innovation programme under the Marie Sk\l{}odowska-Curie grant agreement No.~101030988 and by the
Slovenian Research Agency (ARRS) under the grant QTE (Grant No. N1-0109).

\end{acknowledgments}

\vfill\null

\section*{Supplemental Material\label{sec:methods}}

\subsection*{Experimental realization and measurements}

We realise a pair of strongly tunnelling coupled 1D superfluids by cooling down $^{87}\mathrm{Rb}$ atoms in a double-well potential in an atom chip setup. The initial state is prepared by cooling the atoms directly into a double-well potential. The initial state in this case is a thermal equilibrium state with typical temperatures of $30$-$120\,\si{\nano \K}$ and linear atomic densities of $\approx 70\,\si{\micro\meter^{-1}}$ in a box-like potential with length $\approx 50\,\si{\micro \meter}$.

To achieve lower effective temperatures, we first cool down the atoms in a dressed single-well potential, where the first excited state is in the vacuum state. Slowly splitting the cloud into two parts, maps the ground state and the first excited state to two states with a smaller energy gap and symmetric and anti-symmetric wave functions. The resulting prethermalized state has a lower effective temperature in the anti-symmetric modes (relative degrees of freedom)~\cite{gring2012prethermalization}. This method has been deployed in the measurement with effective temperature $T_\mathrm{eff} = 14\, \si{\nano \K}$ presented in Fig. \ref{fig:s_mi_temp} and \ref{fig:extnd_s_mi}.

To probe the system, we turn off all the traps, and we let the atoms fall freely for $15.6\, \si{\milli \s}$. We measure the projected 2D atomic density distributions via absorption imaging, from which we extract the relative phase between the two condensates for different points along the 1D direction $z$.
Due to the destructive nature of the imaging process, we repeat the measurement hundreds of times to accumulate statistics. Thus, all the expectation values calculated are obtained through ensemble averaging.

\subsection*{Quantum field simulation using coupled quasi-one-dimensional superfluids}

As it has been discussed in several earlier works (see, e.g., Refs.~\cite{exp_sG, Gaussification1, Recurrence}), low energy excitations of coupled parallel one-dimensional gases of weakly interacting atoms can be utilised as quantum field simulator of the sine-Gordon Hamiltonian,
\begin{multline}
    H_{\mathrm{sG}}=\int_0^L \mathrm{d}z\biggl[\god \delta\rho^{2} + \frac{\hbar^{2}\nod}{4m}\left(\partial_{z}\varphi\right)^{2} \\ -2\hbar J \nod \cos(\varphi)\biggr]\, . \label{eq:HsG}
\end{multline}
This model describes the relative phase, $\varphi$, and the relative density fluctuations of two superfluids (see Fig.~\ref{fig:experiment}a). These two fields are canonical conjugate of each other, i.e. $[\varphi(z),\delta\rho(z')]=-\mathrm{i}\delta(z-z')$. In (\ref{eq:HsG}), $m$ is the atomic mass, $\nod$ the uniform atomic density of the condensates, $\god$ the inter-atomic interaction, $J$ the single-particle tunnelling rate.

For sufficiently cold gases in the strong coupling regime, i.e., when the phase coherence length
\begin{align}
    \lambda_{T} =\frac{2\hbar^{2}\nod}{mk_{B}T}
\end{align}
is larger than the healing length of the relative phase (correlation length)
\begin{align}
    l_{C}=\sqrt{\frac{\hbar}{4mJ}}\ ,
\end{align} 
the cosine term in (\ref{eq:HsG}) can be expanded to second order and be approximated by the quadratic Klein-Gordon model.
Introducing the sound velocity $c$, Luttinger parameter $K$ and the Klein-Gordon quasi-particle mass, $M$, given in
terms of the microscopic parameters by
\begin{align}
    c & =\sqrt{\frac{\god \nod}{m}}
    \label{appv}\ ,\\
    K & =\frac{\hbar\pi}{2}\sqrt{\frac{\nod}{m\god}}
    \label{appK}\ , \\
    M & =2m\sqrt{\frac{\hbar J}{\god \nod}}\ ,
\end{align}
the Klein-Gordon (KG) Hamiltonian can be written as
\begin{multline}
    H_{\mathrm{KG}}= \frac{\hbar c}{2}
    \int_0^L\mathrm{d}z\,\biggl[\frac{\pi}{K}\delta\rho^{2}(z) + \frac{K}{\pi}\left(\partial_{z}\varphi(z)\right)^{2}\biggr] \\ + \frac{M^2 c^4}{2\hbar c}
    \int_0^L\mathrm{d}z\;\frac{K}{\pi}\varphi^{2}(z)\ .
    \label{H_KG}
\end{multline}
Note that the first two terms in (\ref{H_KG}) are the Tomonaga-Luttinger liquid (TLL) Hamiltonian. In table~1, the relevant parameters for each measurement are listed, where $\omega_M = Mc^2/\hbar$ is the KG mass in units of angular frequency.

\subsection*{Reconstruction of the initial full covariance matrix}

To extract the full covariance matrix that characterises the state of the system, we use the quantum tomography method developed in \cite{Tomography}. Given that only one of the two canonical variables (the phase) is accessible through experimental measurements, its canonically conjugate variable (the density fluctuations) can be accessed indirectly by letting the system evolve under a harmonic Hamiltonian with known mode frequencies and measuring the phase at different times. 
For each of the harmonic modes, the dynamics correspond to a rotation in phase space, so that over time, the initial density variance turns into phase variance and vice versa. Thus, we can fully reconstruct the initial covariance matrix from phase measurements at a sufficiently large number of different times. This reconstruction is done by first going to Fourier space, where the modes evolve independently, and then fitting the data for the time evolution of the phase covariance of each mode to the known functions expressing this data in terms of the initial phase and density correlations. 

However, one mode needs special treatment: The zero mode, i.e. the mode corresponding to zero momentum, therefore zero energy. This mode does not rotate in phase space; instead, it moves at a constant velocity. This means that the zero mode phase variance does not oscillate in time but grows as a quadratic function of time instead, an effect known as phase diffusion \cite{phase_diffusion_1}. Moreover, because of the compactified nature of the phase-field, which means that phases differing by $2\pi$ should be considered identical, even though the phase grows in time with no bound, measurements can only observe its growth within the interval $[-\pi,+\pi]$. Therefore, the estimation of the initial phase and density correlations of the zero mode is different from that of the others. For the zero mode, we fit a quadratic function instead of an oscillatory function of time, and we restrict the fit to times before reaching the upper bound due to compactification.

More specifically, the dynamics are chosen to follow the Tomonaga-Luttinger liquid model with Hamiltonian
\begin{equation}
    H_\mathrm{TLL}=\int_{0}^{L}dz\left[\god(\delrho(z))^{2}+\frac{\hbar^{2}n_{1D}}{4m}\left(\partial_{z}\varphi(z)\right)^{2}\right] \label{H_LL}
\end{equation}

For a hard-wall box trap, the vanishing of the particle current at the edges of the system means that the effective boundary conditions are of Neumann type 
$\partial_{z}\varphi(x=0)=\partial_{z}\varphi(x=L)=0$. In this case, using cosine eigenfunctions
\begin{align} \label{eq:wavefun}
f_{n}^{\varphi}(z) & =\begin{cases}
2\left(n\hbar\pi\sqrt{\frac{n_{1D}}{\god m}}\right)^{-1/2}\cos(n\frac{\pi}{L}z)\,, & n>0\\
1\, , & n=0
\end{cases} \nonumber \\ 
f_{n}^{\delrho}(z) & =\begin{cases}
-\frac{1}{L}\left(n\hbar\pi\sqrt{\frac{n_{1D}}{\god m}}\right)^{1/2}\cos(n\frac{\pi}{L}z)\,, & n>0\\
-\frac{1}{L}\,, & n=0
\end{cases}\,,
\end{align}
the Hamiltonian can be diagonalised in terms of cosine Fourier modes:
\begin{align}
    H & =\frac{\hbar u}{2}\delrho_{0}^{2}+\sum_{n=1}^{\infty}\frac{\hbar\omega_{n}}{2}\left[\delrho_{n}^{2}+\varphi_{n}^{2}\right]\, , \label{eq:H_TLL_modes}
\end{align}

with
\begin{align}
    u & = 2\god / \hbar L\\
    \omega_n & = ck_n \, . \label{eq:wn}
\end{align}
Furthermore:
\begin{align}
    k_n & = n\pi/L \\
    c & = \sqrt{\god \nod/m} \\
    \god & = \hbar \omega_\perp a_\mathrm{s} \frac{2+3 a_\mathrm{s} \nod}{\left( 1 + 2a_\mathrm{s} \nod \right)} \, . \label{eq:god}
\end{align}
Here, $c$ is the speed of sound and $\god$ the density-broadened 1D interaction strength~\cite{schweigler2019thesis}, $a_\mathrm{s} = 5.2\,\si{\nano \meter}$ is the three-dimensional scattering length~\cite{vanKempen2002RbInteraction} and $m = 1.44\times 10^{-25}\,\si{\kg}$ the mass of a $^{87}\mathrm{Rb}$ atom.

As mentioned before and reflected in the TLL Hamiltonian in Fourier space (Eq. \ref{eq:H_TLL_modes}), we have to treat the zero mode separately. We begin with the harmonic part of the Hamiltonian ($n>0$) where the time evolution of the modes is given by
\begin{align}
    \delta\rho_{n}(t) & =\delta\rho_{n}(0)\cos(\omega_nt) + \varphi_{0}(0) \sin(\omega_nt) \label{eq:time_eve_n_I} \\
    \varphi_{n}(t)	&  =\varphi_{0}(0)\cos(\omega_nt) -  \delta\rho_{n}(0) \sin(\omega_nt) \label{eq:time_eve_n_II}\, , 
\end{align}
In the experiment, we use matter interferometry to measure spatially resolved relative phase between two superfluids, from which, a referenced second-order correlation is calculated for each time step:
\begin{equation} 
    \Phi^2_{ab}(t) = \left<\bigl(\varphi(z_a,t)-\varphi(z_0,t)\bigr) \bigl(\varphi(z_b,t)-\varphi(z_0,t)\bigr) \right>\, .
\end{equation}
Note that subtracting the phase of an arbitrary reference position $z_0$ will only remove the zero mode and does not affect any of the higher modes.

Expanding $\Phi^2$ with eigenfunctions (\ref{eq:wavefun}), gives
\begin{equation}
    \Phi^2_{ab}(t)  =  \sum^{N}_{j,k = 1} f^{a,b}_{j,k} \left<\varphi_j(t)\varphi_k(t) \right>
\end{equation}
where
\begin{equation}
    f^{a,b}_{j,k} = \bigl(f^\phi_j(z_a) - f^\phi_j(z_0)\bigr) \bigl(f^\phi_k(z_b) - f^\phi_k(z_0)\bigr)\,.
\end{equation}

Using the equation of motion (\ref{eq:time_eve_n_II}), and defining $\tilde{Q}_{jk} = \left<\varphi_j(0)\varphi_k(0) \right>$, $\tilde{R}_{jk} = \left<\frac{1}{2}\left\{\varphi_j(0),\delrho_k(0)\right\} \right>$, and $\tilde{P}_{jk} = \left<\delrho_j(0)\delrho_k(0) \right>$, we obtain
\begin{equation}\label{eq:Phi_full}
    \begin{split}
        \Phi^2_{ab}(t)  &=  \sum^{N}_{j,k = 1} f^{a,b}_{j,k} \cos(\omega_jt)\cos(\omega_kt) \tilde{Q}_{jk}\\
                         &+  \sum^{N}_{j,k = 1} (-f^{a,b}_{j,k}-f^{a,b}_{j,k}) \cos(\omega_jt)\sin(\omega_kt)\tilde{R}_{jk}\\
                         &+  \sum^{N}_{j,k = 1} f^{a,b}_{j,k} \sin(\omega_jt)\sin(\omega_kt) \tilde{P}_{jk} \, .
    \end{split}
\end{equation}

Equation (\ref{eq:Phi_full}) stands in the heart of the tomography process: The goal is to find the elements of $\tilde{Q}$, $\tilde{R}$, and $\tilde{P}$ using an optimization process. Note that the left hand side is calculated using measured phase profiles in the experiment and in the right hand side, $f^{a,b}_{j,k}$ and $\omega_j$ can be calculated from the experimental parameters, as shown in (\ref{eq:wavefun}) and (\ref{eq:wn}) respectively. Results are presented in SM. For a more detailed explanation of the tomography process, please refer to~\cite{Tomography}.

Unlike all higher modes, the zero mode, which corresponds to the constant in space eigenfunction, is not a harmonic oscillator mode of the Tomonaga-Luttinger liquid Hamiltonian (equation (\ref{H_LL})). This is because only one of the canonical variables, $\delta\rho_{0}$, is present in the Hamiltonian for the zero mode. As a result, the time evolution of the zero mode is given by
\begin{align}
    \delta\rho_{0}(t) & =\delta\rho_{0}(0)=\mathrm{const}, \\
    \varphi_{0}(t)	&  =-u\delta\rho_{0}(0)t+\varphi_{0}(0)\, , 
\end{align}
which means that the phase variance grows with time as 
\begin{equation}
    \left\langle \varphi_{0}(t)^{2}\right\rangle = \left\langle \varphi_{0}^{2}\right\rangle _{t=0}-\left\langle \left\{ \varphi_{0},\delta\rho_{0}\right\} \right\rangle _{t=0}ut+\left\langle \delta\rho_{0}^{2}\right\rangle _{t=0}u^{2}t^{2}
\end{equation}

However, because of the compactified nature of the phase-field, its zero mode component $\varphi_{0}$ is not a well-defined, measurable operator. Only imaginary exponentials of the form $e^{{\rm i}n\varphi_{0}}$ for integer $n$ are well-defined. Nevertheless, under the assumption the initial state is Gaussian in terms of the zero mode (as also for all other modes too) and given that it remains Gaussian under the dynamics following from $H_\mathrm{TLL}$, we can derive the zero mode variance from the mean value of $e^{{\rm i}n\varphi_{0}}$ using the cumulant expansion formula for the special case of Gaussian random variables
\begin{equation}
    \left\langle \exp\left(\mathrm{\mathrm{i}}\varphi_{0}\right)\right\rangle =\exp\left(\mathrm{\mathrm{i}}\left\langle \varphi_{0}\right\rangle -\left\langle \varphi_{0}^{2}\right\rangle /2\right)
    \label{eq:cumulant}
\end{equation}
From the above, we find
\begin{equation}
    \left\langle \varphi_{0}^{2}\right\rangle =-2\log\lvert \left\langle \exp\left(\mathrm{\mathrm{i}}\varphi_{0}\right)\right\rangle \rvert    
    \label{eq:phase_diffusion}
\end{equation}

Therefore, in order to extract the zero mode part of the covariance matrix $\left\langle \varphi_{0}^{2}\right\rangle _{t=0},\left\langle \delta\rho_{0}^{2}\right\rangle _{t=0}$ and $\left\langle \left\{ \varphi_{0},\delta\rho_{0}\right\} \right\rangle _{t=0}$ in the initial state we calculate the zero mode variance of the phase at each time from (\ref{eq:cumulant}) and fit this with the theoretical formula (\ref{eq:phase_diffusion}).

Having the covariance matrix in the Fourier space for the first $N$ modes, we use a discrete Fourier transformation based on the eigenfunctions (equation (\ref{eq:wavefun})) to calculate the covariance matrix $\Gamma$ in the real space. We chose the cut-off based on the reconstructed occupation numbers. We only take into account modes with physical (positive) occupation numbers.
This covariance matrix is used to calculate vN entropy and MI, as discussed in the main text and Fig.  \ref{fig:experiment}C.

\subsection*{Covariance matrix of Klein-Gordon model in thermal equilibrium}

The theory predictions in our work are calculated based on the covariance matrix of the thermal equilibrium states of Klein-Gordon (KG) model (equation \ref{eq:KG}), which is given by~\cite{lebellac1996thermalfieldtheory}:
\begin{align}
    Q_{ij} &= \frac{\pi}{2KL}\frac{\hbar c}{Mc^{2}}\coth\left(\frac{Mc^{2}}{2k_{B}T}\right)\nonumber  \\   & + \frac{\pi}{KL}\sum_{n=1}^{N}\frac{\hbar c}{\epsilon_{n}}\coth\left(\frac{\epsilon_{n}}{2k_{B}T}\right)\cos(k_n z_i)\cos(k_n z_j) \\
    R_{ij} &= 0 & \\
    P_{ij} &= \frac{K}{2\pi L}\frac{Mc^{2}}{\hbar c}\coth\left(\frac{Mc^{2}}{2k_{B}T}\right)\nonumber \\  + & \frac{K}{\pi L}\sum_{n=1}^{N}\frac{\epsilon_{n}}{\hbar c}\coth\left(\frac{\epsilon_{n}}{2k_{B}T}\right)\cos(k_n z_i)\cos(k_n z_j) \, , \label{eqSup:KGthermal}
\end{align}
with the dispersion relation $\epsilon_{n}=\sqrt{\hbar^{2} k_n^{2}c^{2}+M^{2}c^{4}}$. Here, $M$ is the KG mass, $L$ the system size, $K$ the Luttinger parameter, $T$ the temperature, $N$ the UV cut-off and the rest of the parameters are defined before. The next section will explain how $M$ and $T$ are estimated based on the measured data.
To include the effect of the finite imaging resolution, we convolve the theoretical calculations with a Gaussian point-spread function with a standard deviation $\sigma_\mathrm{PSF}\approx 3\, \si{\micro \meter}$~\cite{schweigler2019thesis} (see next section).

\subsection*{Estimation of the temperature and Klein-Gordon mass}

In order to compute theoretical predictions for the mutual information in the initial state based on the assumption that these are the thermal states of the KG model, we need to estimate two effective parameters, the mass and the temperature. We do this by fitting the results of the tomographic reconstruction for the mode variances to those corresponding to KG thermal states. Given that the modes are decoupled from each other both initially and throughout the tomography dynamics, the estimation of the KG mode frequency can be done independently for each mode. Having estimated the mode frequencies, we can then verify if they follow the theoretical dispersion relation of the KG model and extract the corresponding mass parameter by a fit. 

The relation between the post-quench quadratures and the initial (pre-quench) KG state with mode occupation number $N_{0n}$ is given by
\begin{align}
\left\langle \varphi_n^2 \right\rangle  & =\frac{\omega_{n}}{\omega_{0n}}\left( N_{0n}+\frac{1}{2}\right) \label{eq:phin2}\\
\left\langle \delrho_n^2 \right\rangle  & =\frac{\omega_{0n}}{\omega_{n}}\left( N_{0n}+\frac{1}{2}\right)\label{eq:rhon2}
\end{align}
where $\omega_{0n}$ and $\omega_{n}$ are the pre- and post-quench mode frequencies respectively. Both $\left\langle \varphi_n^2 \right\rangle $ and $\left\langle \delrho_n^2 \right\rangle $ are achieved via tomography and $\omega_n$ is given by the equation (\ref{eq:wn}). From the equations (\ref{eq:phin2}) and (\ref{eq:rhon2}) we can calculate both $\omega_{0n}$ and $N_{0n}$:
\begin{align}
N_{0n}  & = \sqrt{\left\langle \varphi_n^2 \right\rangle \left\langle \delrho_n^2 \right\rangle } - \frac{1}{2} \label{eq:nk}\\
\omega_{0n}  & = \omega_n \sqrt{ \frac{\left\langle \delrho_n^2 \right\rangle}{\left\langle \varphi_n^2 \right\rangle}}\, . \label{eq:wk}
\end{align}
Assuming the initial state is thermal, we use the following fit function to extract the temperature, $T=(\beta k_\mathrm{B})^{-1}$:
\begin{align}
    N_{0n}^\mathrm{fit} & = \mathrm{exp}\left( {-k_n^2 \sigma_\mathrm{PSF}^2/2} \right) (\frac{1}{\mathrm{exp}(\hbar \omega_{0n} \beta) - 1} + \frac{1}{2}) - \frac{1}{2}\, .
\end{align}
Note that we also introduce the effect of the imaging system by multiplying the modes with a Gaussian point spread function with width $\sigma_\mathrm{PSF}$ which corresponds to a convolution with a Gaussian point-spread function in the real space~\cite{schweigler2019thesis}.
To extract the KG mass, we fit the KG dispersion relation,
\begin{equation}
\omega_{0n}^\mathrm{fit} = \sqrt{c^2 k_n^2 + M^2 c^4/\hbar^2}\, ,
\end{equation}
to the calculated $\omega_{0n}$.

\subsection*{Tomography results}

\begin{figure*}
    \centering
    \includegraphics[width=0.8 \textwidth]{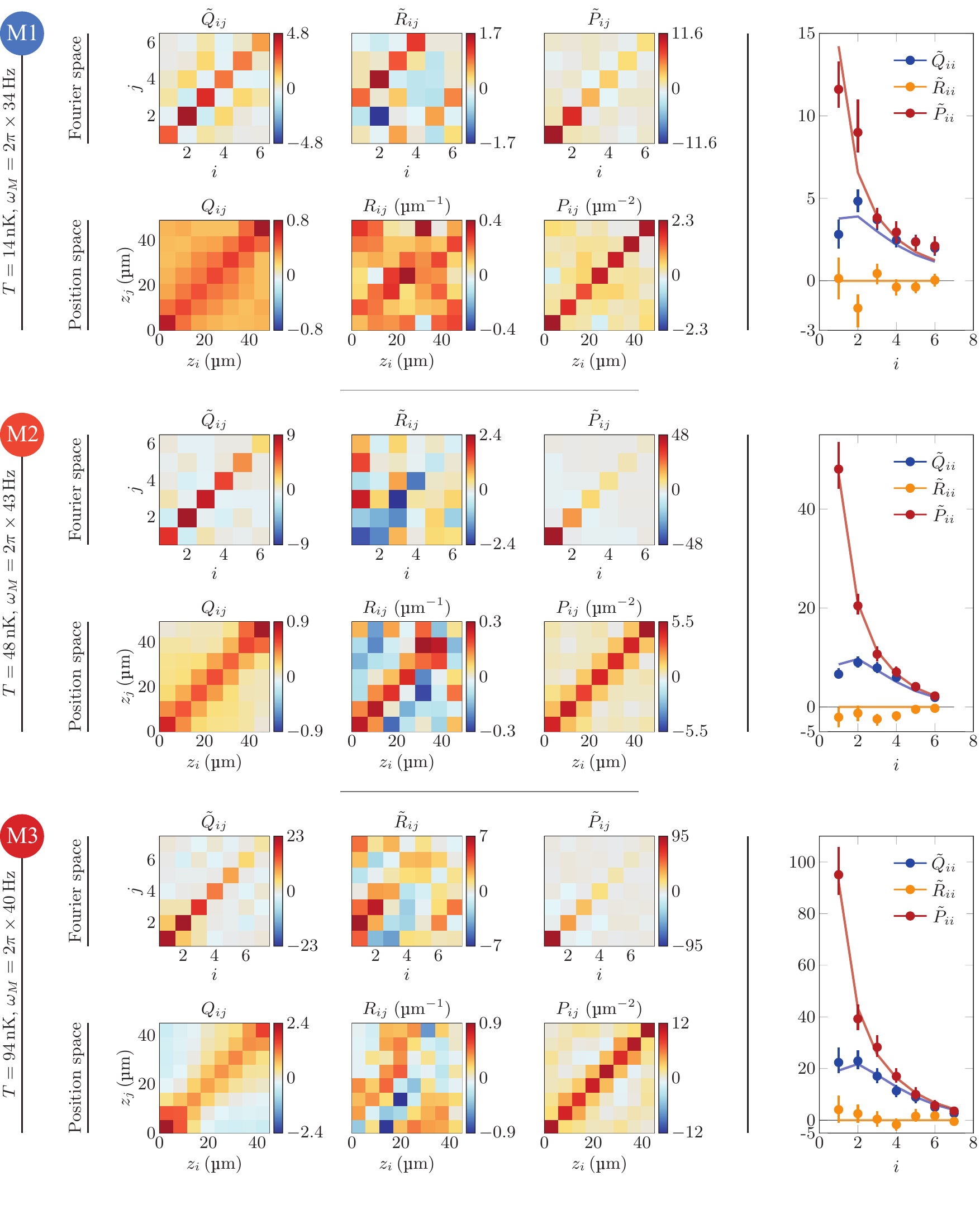}
    \caption{\textbf{Tomography results for three measurements from lowest to highest temperature (M1 to M3).} For each measurement, the elements of the covariance matrix in Fourier and position space are presented. On the right, the diagonal elements of the covariance matrix in the Fourier space are plotted. The error bars represent 95\% confidence intervals obtained via bootstrapping. Solid lines are the theory calculations corresponding to the extracted temperature and KG mass for each measurement.}
    \label{fig:QRP_all}
\end{figure*}

\begin{table*}
\begin{center}
\begin{tabular}{ c c c c c c c}
\hline
 Measurement   & $T\, (\si{nK})$ & $\lambda_T \, (\si{\micro\meter})$ & $\omega_M/2\pi\, (\si{\Hz})$ & $l_\mathrm{C}\, (\si{\micro\meter})$ & $q = \lambda_T / \ell_{J} $ & $r = \hbar \omega_M / k_\mathrm{B}T$ \\ 
 \hline
M1   & 14 & 37 & 34 & 7.3 & 5 & 0.11\\  
M2   & 48 & 18 & 43 & 6.8 & 2.6 & 0.043\\
M3   & 94 & 9.5 & 40 & 7.4 & 1.3 & 0.021\\  

\hline
\end{tabular} 
\end{center}

\caption{\textbf{List of relevant parameters for the three measurements.} $T$ is the temperature, $\lambda_T$ the thermal coherence length, $\omega_M=Mc^2/\hbar$ the KG mass in angular frequency units, and $l_{C}$ the healing length of the phase. \label{table:exppar}}
\end{table*}

\begin{figure*}
    \centering
    \includegraphics[width=0.8\textwidth]{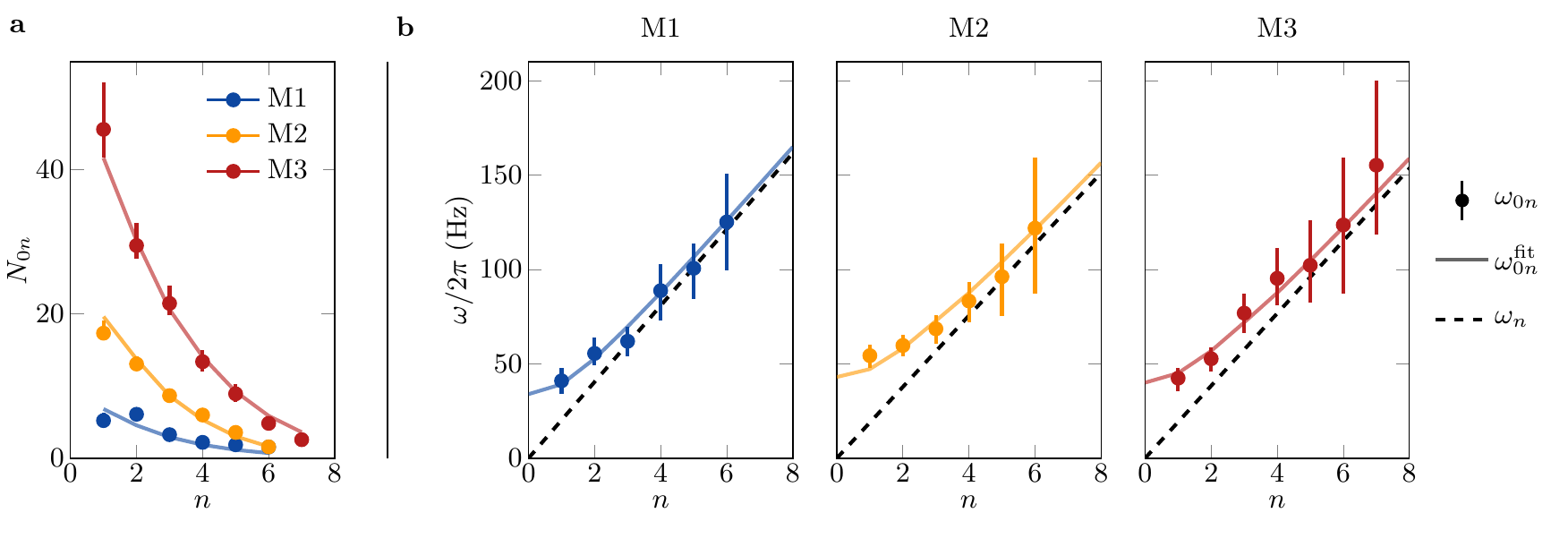}
    \caption{\textbf{Extraction of temperature and KG mass from the mode occupation number and dispersion relation} \textbf{a,} For each measurement $N_{0n}$ is plotted along with a fit curve (solid line), $N_{0n}^\mathrm{fit}$. \textbf{b,} Dispersion relation for three different measurements are plotted. For every quantity, the error bars represent 95\% confidence intervals obtained via bootstrapping.}
    \label{fig:n_w_all}
\end{figure*}

The results of the reconstruction process for the three measurements are presented in Fig.~\ref{fig:QRP_all} and~\ref{fig:n_w_all} and the corresponding parameters are written in Table~\ref{table:exppar}. We show the covariance matrix elements obtained after the tomography in the Fourier space, as well as the elements in the position space after a discrete inverse Fourier transform. Note that the only constraint on the elements of $\tilde{\Gamma}$ is that the resulting state is physical i.e. the Heisenberg uncertainty relation holds~\cite{Tomography}. We do not restrict the state by any further assumption. For example, we do not assume that the state is Gaussian (see next section) or $\tilde{R}$ is small. The only assumption in the tomography process is that the post-quench dynamics follow the Tomonaga-Luttinger liquid Hamiltonian for short times as mentioned before.

\subsection*{Fourth order correlation}
\begin{figure}[h]
    \centering
    \includegraphics{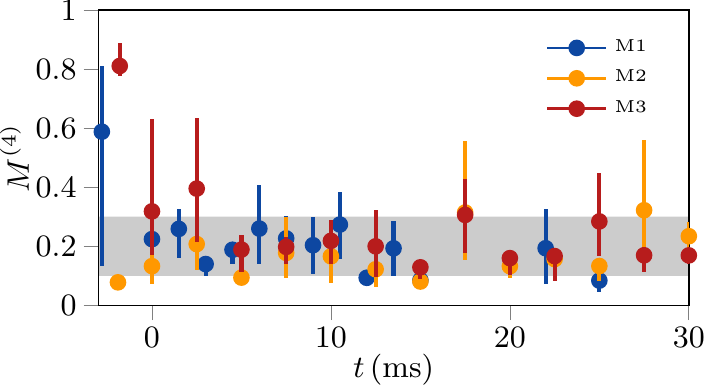}
    \caption{\textbf{Fourth order correlations} The error bars represent 95\% confidence intervals obtained via bootstrapping. The shaded area is the typically calculated bias~\cite{Gaussification2}.}
    \label{fig:M4_all}
\end{figure}

In order to show that the reconstructed state is Gaussian, we measure the normalised averaged connected fourth-order correlation function, $M^{(4)}$. Please refer to previous extensive works by our group~\cite{Gaussification1, Gaussification2, schweigler2019thesis}. The results for the four measurements are shown in Fig.~\ref{fig:M4_all}.

\subsection*{Effect of finite imaging resolution}

\begin{figure}[b]
    \centering
    \includegraphics{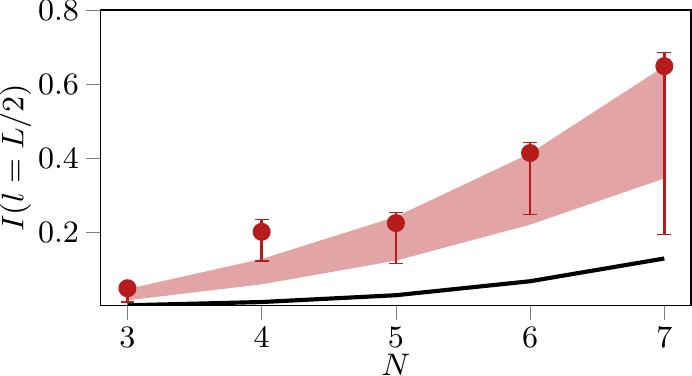}
    \caption{\textbf{Imaging effect on mutual information.} Mutual information for $l=L/2$ for different cut-off, $N$. The error bars represent 95\% confidence interval. The shaded area is the theoretical calculation and the solid black line is the residual MI due to the effect of finite imaging resolution.}
    \label{fig:mi_sigma_eff}
\end{figure}

\begin{figure*}
    \centering
    \includegraphics[width=0.9\textwidth]{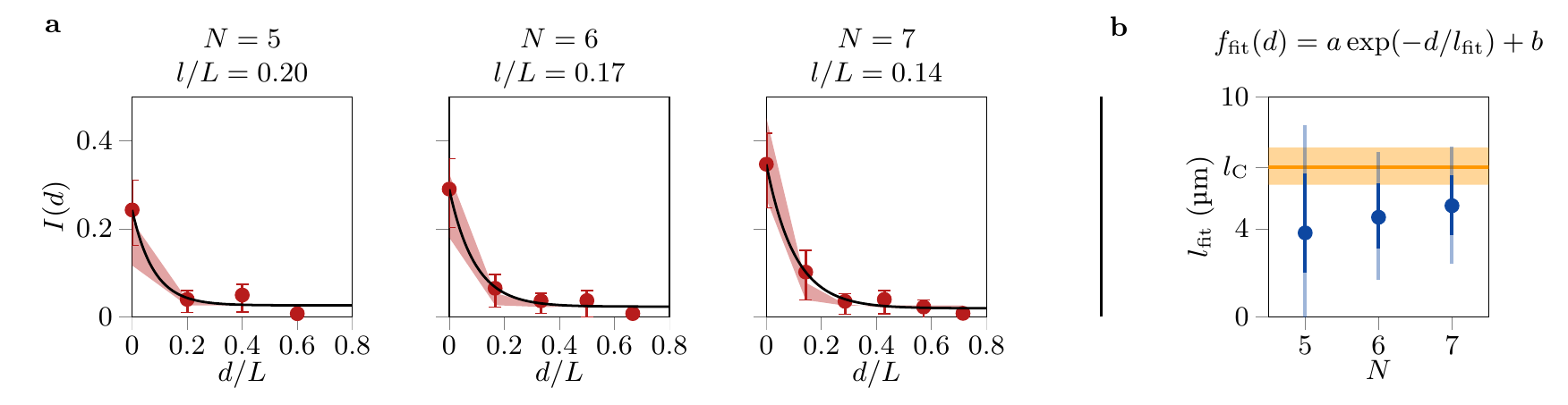}
    \caption{\textbf{Dependence of $l_\mathrm{fit}$ on the total number of modes, $N$.} \textbf{a,} For different $N$, the same quantities are plotted as in Fig.~3 in the main text but here, we shifted the theory calculation by $b$, to represent better the agreement between theory, experimental data and the fit. Note that changing the total number of modes consequently changes the size of the sub-region. \textbf{b,} The resulting decay length, $l_\mathrm{fit}$ is plotted for different $N$. the solid part of the error bars represents 68\% confidence interval (standard error) and the extended transparent part 95\% confidence interval. The solid orange line and the shaded area show the $l_\mathrm{C} = 6.8 (6.0, 7.7)$ and its 95\% confidence interval. }
    \label{fig:mi_gap_n}
\end{figure*}

We investigate the effect of the finite imaging resolution and the ultra-violet cut-off on the measurement of mutual information (MI). 
The Gaussian point-spread function of the imaging system acts as a soft cut-off. This cut-off not only limits the number of observable modes but also affects the lower modes that can be measured~\cite{schweigler2019thesis}. In length scales close to and smaller than the imaging resolution, the blurring effect creates correlations, producing unphysical MI. We theoretically quantify this "fake" MI by considering a state with high temperature, and short correlation length (much shorter than the imaging resolution), whose MI is practically zero ($\sim 10^{-6}$). For this state, we calculate the MI of the half-system, $l=L/2$, for different numbers of modes, $N$. The result is presented in Fig.~\ref{fig:mi_sigma_eff} with solid black lines along with the experimental data. The error bars are $95\%$ confidence intervals obtained via bootstrapping and the shaded area is the $95\%$ confidence interval for the theoretical calculation. Note that the calculated residual MI only depends on the imaging resolution. The figure suggests that our measured MI is significantly higher than what is calculated to be the residual MI due to the imaging effect. 

We further investigate the effect of the hard cut-off on the decay of MI as discussed in the main text (Fig. 3).
In Fig.~\ref{fig:mi_gap_n}\textbf{a}, for each $N$, the information similar to Fig.~2 in the main text is shown. 
In Fig.~\ref{fig:mi_gap_n}\textbf{b}, the dependence of $l_\mathrm{fit}$ on the total number of modes, $N$, is plotted. We observe that for all values of $N$, experimental data and the theoretical calculations are in good agreement. As the number of modes decreases, sub-region size and spacing increase, which lowers the quality of fit, resulting in large error bars. Nevertheless, the decay length stays consistent for different cut-offs. Note that for $N<5$ the sub-regions become too large and reliable fitting is no longer possible. 

\newpage
\providecommand{\noopsort}[1]{}\providecommand{\singleletter}[1]{#1}%

\end{document}